\documentstyle[aps,multicol,epsf]{revtex}
\newcommand{\beq}{\begin{equation}}
\newcommand{\eeq}{\end{equation}}
\newcommand{\bdis}{\begin{displaymath}}
\newcommand{\edis}{\end{displaymath}}
\newcommand{\bea}{\begin{eqnarray}}
\newcommand{\eea}{\end{eqnarray}}
\newcommand{\barr}{\begin{array}}
\newcommand{\earr}{\end{array}}
\begin{document}
\title{Sandpile Model on Sierpinski Gasket Fractal}
\author{Brigita Kutjnak-Urbanc$^1$, Stefano Zapperi$^1$, 
Sava Milo\v sevi\'{c}$^2$,
H. Eugene Stanley$^1$}
\address{$^1$ Center for Polymer Studies and Department of Physics, Boston
University, Boston, Massachusetts 02215\\
$^2$ Faculty of Physics, University of Belgrade, P. O. Box 368,
11001 Belgrade, Yugoslavia.} 
 
\date{lc6136.tex -- \today{} draft}
\maketitle
 
\begin{abstract}
We investigate the sandpile model on the two--dimensional
Sierpinski gasket fractal. We find that the model 
displays novel critical behavior, and we analyze the
distribution functions of avalanche sizes, lifetimes and
topplings and calculate the associated critical exponents
$\tau = 1.51 \pm 0.04$, $\alpha = 1.63 \pm 0.04$ and $\mu 
= 1.36 \pm 0.04$. The avalanche size distribution shows
power law behavior modulated by logarithmic oscillations
which can be related to the discrete scale invariance of
the underlying lattice. Such a distribution can be formally
described by introducing a complex scaling exponent
${\tau}^{*} \equiv \tau + i \delta$, where the real part
$\tau$ corresponds to the power law and the imaginary part
$\delta$ is related to the period of the logarithmic
oscillations.
 
\end{abstract}
 
\pacs{PACS numbers: }
 
\begin{multicols}{2}
 
\narrowtext
 
\section{Introduction}
 
The concept of self--organized criticality (SOC) has been
introduced by Bak et al. \cite{soc} to describe the
tendency of a large class of dynamical systems to
spontaneously evolve into a critical state without fine
tuning of any external parameter.  Sandpile models
\cite{soc,zhang} have been introduced as an example of this
kind of phenomena and have been widely studied numerically
and analytically \cite{nagel,manna,priezzhev,ola,grass}.
Two principal analytical approaches have been followed: the
first involves the group theory formalism introduced by
Dhar \cite{dhar} and the second is a real space
renormalization scheme recently developed by Pietronero et
al.\cite{zap}. Other theoretical approaches involve
nonlinear continuous differential equations
\cite{diaz,hwa}.
 
Sandpile models are inspired by the dynamics of sand
flowing along the slope of a pile. By adding sand grains to
the pile the system eventually reaches a stationary state
characterized by avalanches of all length scales. The term
criticality refers here to the absence of any
characteristic length scale in this state.  Sandpile models
have been studied mostly on Euclidean lattices. It has been
shown that different kinds of Euclidean lattices do not
affect the critical exponents \cite{manna}. This fact is
similar to the universality observed in ordinary critical
phenomena. Moreover, in the case of the Bethe lattice one
recovers the mean field results \cite{dhar2,mf}. 
However, sandpile models, to our knowledge, have not been 
studied on a fractal substrate, in particular on a simple 
deterministic fractal such as that epitomized by
the Sierpinski gasket (SG). 
 
It has been shown,
via specific calculations \cite{ising} and through general 
rigorous analysis \cite{saw}, that for the standard
Ising model (and for some more general models) on finitely
ramified fractals no spontaneous magnetization can
exist at any finite temperature. It might have happened
that, by some assumed analogy, no self-organized critical
behavior has been expected so far to occur on the
deterministic fractals. However, we shall demonstrate that 
the SOC phenomenon exists on the SG fractal and displays 
novel features. Specifically, we study numerically the 
critical height sandpile model on the SG lattice with the 
generator scaling base $b = 2$ which corresponds to the 
fractal dimension $D =\log 3 / \log 2 \approx 1.58$.  
We calculate the distribution of avalanche sizes, their 
lifetimes, and topplings. The avalanche size distribution shows 
a power law behavior modulated by logarithmic oscillations. 
This kind of oscillation has been already observed in the
scaling functions of different systems
\cite{peak}, and here it can be related to the 
discrete scale invariant nature of the underlying fractal
lattice. It is interesting to note that complex scaling exponents
have been recently detected in 
earthquakes statistics \cite{complex}.
 
The measured scaling exponents vary with the system size $L$ and 
the values, extrapolated to $L \to \infty$, appear to differ from 
those computed on the Euclidean lattices. Computing expectation 
values, we are able to verify the relationships between different 
critical exponents.
 
In addition, we investigate time correlations of the number of
drops and topplings during the avalanche. Calculating the power 
spectra, we find that as in the case of the two--dimensional 
Euclidean lattice \cite{kk,jensen} there are no long--range 
temporal correlations.
 
\section{The model}
 
Our cellular automata model is defined on the SG lattice as
shown in Fig.~1. The number $n$ is related to the number of 
sites $L = 2^n + 1$ along one direction of the lattice and is 
used hereafter as a measure of the system size. Within the 
sandpile model, all the sites of the fractal lattice are exposed 
to the same local dynamical rules. The exceptions   
are the three apex sites where the sand grains flow out
of the system. The dynamics begins when we
associate an integer height variable $z_i$ with every
lattice site $i$. At each later step one lattice site is chosen
at random and its height is increased by one. Whenever the
height on a site $i$ reaches the critical value $z_c=4$,
the site becomes unstable (active) and relaxes according to the
following rules:
\bea
z_i\rightarrow z_i-4 , \\
z_j\rightarrow z_j+1 ,
\eea
where $j$ are the nearest neighbors of $i$. These rules
conserve the total number of grains, except on the three
apex sites (independent of the system size) where two sand
grains are lost. Successive relaxation events generate the
sand flow that eventually brings the sand out of the
system. Due to the local conservation, imposed by the
dynamical rules, the system finally evolves into the
stationary state characterized by the balance between the
input and the output flow.
 
The critical exponents are extracted from avalanche
distributions in the stationary state.  
We define the size $S$ as the number of
distinct sites visited by an avalanche, the toppling size
$m$ as the number of relaxation events and the lifetime $T$
as the number of updating steps during an avalanche.  All
these quantities are expected to be distributed as power
laws
\bea 
P(S) \sim S^{-\tau} , \\
P(m) \sim m^{-\mu} , \\
P(T) \sim T^{-\alpha} , 
\eea
where $\tau$, $\mu$, and $\alpha$, are critical exponents of the 
respective distribution functions $P(S)$, $P(m)$ and $P(T)$. We can 
relate these exponents by considering conditional expectation values 
for an average avalanche size $<S_T>$ and an average toppling size 
$<m_T>$ at a given avalanche lifetime $T$:
\beq
<S_T>\sim T^{\beta} \, ,
\label{beta}
\eeq
\beq
<m_T>\sim T^{\gamma} \, ,
\label{gamma}
\eeq
and similar other relations. By taking into account the
definitions of critical exponents, given by Eqs.(3)-(7), 
scaling relations between exponents can be derived \cite{ola}:
\beq
\tau=1+(\alpha-1)/\beta \, , 
\eeq
\beq
\mu=1+(\alpha-1)/\gamma \, .
\eeq
 
Finally we study temporal correlations  by considering 
the number of active sites and the number of grains
falling out of the system at each time step.
The power spectrum of this signal falls off as
\beq
S(f)\sim f^{-\phi}
\eeq
In the Euclidean case, $\phi=2$ showing the absence of
long--range temporal correlations.
 
\section{Simulation results}
 
We perform numerical simulations for different lattice
sizes ranging from $n=3$ to $n=7$. The total number of
sites $S_{n+1}$ of the system size $n + 1$ is related to
the number of sites $S_n$ of the system size $n$ via the
equation $S_{n + 1} = 3 S_n - 3$ with $S_0 = 3$, which
corresponds to a total number of lattice sites going from
$S_3 = 42$ to $S_7 = 3282$.  
 
A simple way to characterize the properties of the stationary state 
is to compute the fraction $p_z$ of sites having height $z_i=1$, $2$, 
$3$ and $4$. We report these results in Table~1 for different system 
sizes, together with an average height $<z>$.
The obtained values are very close to those found on the 
Euclidean lattice \cite{manna,priezzhev}.
 
In Fig.~2 we show the avalanche size distribution for different system 
sizes. One can see that there are quite a few avalanches (represented 
by the last peak of each distribution curve) which span the entire 
lattice. This occurs because the balance between incoming and outgoing 
particles forces the avalanches to reach the three apex sites.
Due to the self--similarity of the underlying lattice, the same 
phenomenon occurs on all fractal substructures, which is manifested 
by a series of peaks on each distribution curve. 
 
The phenomenon described above is reflected in a peculiar behavior
of the active sites during the evolution of an avalanche: the active 
sites are localized (trapped) within fractal substructures for many 
time steps.  Such a trapping does
not occur in the Euclidean lattice where the active sites
are essentially on the avalanche front. This is apparent from 
Fig.~3 where the active sites for a typical avalanche in the Euclidean
square lattice are compared with an avalanche on the fractal lattice.
Similar differences which spring from differences in the topology 
of the lattices were noted before \cite{B2} in a study of
linear polymers on the diamond hierarchical lattice.
 
The power--law behavior in a double logarithmic plot is
modulated by oscillations with a period $p$ that can be
related to the scaling properties of the SG lattice. A
self--similar lattice is left invariant only by a discrete
set of scale transformations, namely by those with a
scaling parameter of the form $\lambda=b^{n}$. Under this
condition, it has been shown \cite{peak2} that the most
general scale invariant function of the real space
coordinates is a power law multiplied by a logarithmically
periodic function. These oscillations can be formally
described by introducing a complex scaling exponent
${\tau}^{*} \equiv \tau + i \delta$ where the real part $\tau$
corresponds to the power law exponent, while the imaginary
part $\delta$ is related to the period of oscillations.
In our case $\delta=2 \pi /p = 2 \pi /\log 3 = 5.72$.  
To extract $\tau$, we fit the distribution with a 
power law modulated by a periodic function.
 
The last peak in the distribution of avalanche sizes is a 
consequence of the fact that at any system size there are only 
three boundary points where the sand can flow out of the system, 
in contrast to the Euclidean lattice where the number of boundary 
points increase in proportion to the system size. However, one can 
study the effect of boundaries by calculating the avalanche size 
distribution of the same sandpile model on the Euclidean lattice 
with only four boundary points, e.g. the four corner sites of the 
square lattice (for the rest of the edge points periodic boundary
conditions apply). The results are shown in Fig.~4 for system
sizes $L = 4$, $8$, $16$ and $32$. The distributions are power 
laws with peaks at the total numbers of sites on the lattice.
 
In the SG case, we report in Figs.~5 and 6 the
distributions of avalanche lifetimes and topplings in a
double logarithmic plot.  Both distributions display pure
power--law behavior without any modulations. The power--law
regimes grow with the system size.
 
As in the case of the Euclidean lattice
\cite{manna}, the scaling exponents depend on the
system size. We can, however, extract the asymptotic results
by plotting the logarithm of the exponents versus $1/\log L$,
where $L$ is the linear size of the lattice. This relationship 
is presented in Fig.~7, where we depict also the extrapolated 
critical exponents in the limit $n \rightarrow \infty$. We found 
the following extrapolated values ${\tau}_{\infty} = 1.51\pm 0.04$, 
${\alpha}_{\infty} = 1.64\pm 0.04$ and $\mu_{\infty} = 1.36\pm 0.04$
for the avalanche size, lifetime and toppling distributions,
respectively. The distribution functions presented in these figures 
have been calculated by averaging over $2^{17}$ avalanches.
 
To check the consistency of our results we compute the scaling 
exponents of the conditional expectation values, i.e. exponents 
related to the average avalanche size $<S_T>$ and number of topplings 
$<m_T>$ in dependence on the lifetime $T$. Fig.~8 shows 
results of our computation
in a double logarithmic plot. The slopes in the figure correspond
to the exponents $\beta$ and $\gamma$ as defined in Eqs.(\ref{beta}) 
and (\ref{gamma}) and are found to be $\beta = 1.13 \pm 0.05$ and 
$\gamma = 1.73 \pm 0.05$. These two values can be compared with the
ones evaluated from the scaling relations given by Eqs.(8) and (9), 
and using the estimated critical exponents $\tau_\infty$, 
$\alpha_\infty$ and $\mu_\infty$. Thus, we find the values
$\beta = 1.24 \pm 0.12$ and $\gamma = 1.75 \pm 0.18$, which
are in agreement, within the numerical error, with the
directly obtained values from Fig.~8.
 
Finally, within the scope of the sandpile model \cite{soc},
we calculate the temporal correlations of two
quantities: the number of particles which fall out of a
system in a unit time and the number of topplings. The unit
time in this case corresponds to one updating step of the
lattice variables. We calculate the power
spectra of the above two quantities. In both cases we find
a $1/f^2$ type of spectrum. Thereby, the type of temporal
correlations turns out to be the same as in the original
model on a two--dimensional square lattice
\cite{kk,jensen}. Our results are presented in Fig.~9. The
flattening of the $1/f^2$ spectrum at small frequencies is
due to finite size effects. In contrast to other scaling
exponents presented in this paper, the exponents of power
spectra do not vary with the system size, that is, the
$1/f^2$ type of spectrum corresponds to an exponential
decay of temporal correlations independently of the system
size.
 
\section{Conclusion}

We computed numerically the scaling exponents for the avalanche 
distributions in the critical height sandpile model on the Sierpinski 
gasket lattice with the generator base $b = 2$. The lattice coordination 
number is the same as in the two--dimensional square lattice and 
therefore the dynamical rules of the sandpile model are exactly the same. 
The boundaries, however, are different, since on the SG lattice
the sand can flow out only through three sites at every
scale. This fact changes substantially the avalanche
dynamics. The active sites become trapped (localized) and 
topple more than once during a single avalanche.
 
In relation to the standard critical phenomena it is interesting 
to note how the dimensionality and the topology of the lattice 
affect the critical behavior of the model. In one dimension, the 
critical height sandpile model is trivial in that avalanches are not 
power--law distributed \cite{soc}. A similar behavior occurs for 
example in the Ising model where no phase transition is observed 
in dimension less than two. We have shown, however, that on a 
finitely ramified fractal, the sandpile has nontrivial critical 
behavior, in contrast to the Ising model which has no phase 
transition on such a fractal lattice \cite{ising,saw}.
 
Finally, we note that self--similar lattices have been proven very
helpful in constructing exact real space renormalization group
transformations \cite{dhar3,B1} for standard critical phenomena. 
Having demonstrated that self--organized criticality can exist on 
a fractal lattice, it would be beneficial to find such a transformation 
for this model, trying to link the rigorous approach of Dhar et al. 
\cite{dhar} with the real--space renormalization scheme presented 
in \cite{zap,ivash}. 
 
This work was supported by the National Science Foundation. 
B. K.-U. acknowledges additional financial support from 
the Ministry of Science of Slovenia.  We would like to thank 
S. Havlin, K. B. Lauritsen, H. Leschhorn and  L. Pietronero
for useful discussions and comments.

\begin{figure}
\caption{
An example of the SG lattice with the generator $b = 2$, at 
the stage of construction $n=2$, and with linear size $L=5$. 
Each site has four neighbors except for the three apex sites 
with only two neighboring sites. The arrows indicate the 
direction of sand flow from a chosen site.}
\label{fig1}
\end{figure}
 
\begin{figure}
\caption{
The distribution of avalanche sizes of the sandpile model on a SG 
lattice. Different curves correspond to different system sizes. 
The arrows indicate the peaks in the distribution.}
\label{fig2}
\end{figure}
 
\begin{figure}
\caption{A snapshot of an avalanche on the Sierpinski
gasket lattice (a) compared with an avalanche on the
Euclidean square lattice (b). Active sites are depicted in
black and sites that have toppled at least once are colored
in gray.}
\label{fig3}
\end{figure}
 
\begin{figure}
\caption{The distributions of avalanche sizes for a sandpile 
model on a square lattice with four exiting points are depicted 
for several system sizes $L = 4$, $8$, $16$ and $32$. Data for
$L = 16$ and $L = 32$ are averaged at sizes $S > 50$.}
\label{fig4}
\end{figure}
 
\begin{figure}
\caption{The distribution of avalanche lifetimes of the sandpile
model on a SG lattice as calculated for different system sizes.
Data are logarithmically binned at lifetimes $T > 50$.}
\label{fig5}
\end{figure}
 
\begin{figure}
\caption{The distribution of the number of topplings per
avalanche of the sandpile model on a SG lattice
as calculated for different system sizes. 
Data are logarithmically binned at toppling sizes $m > 50$.}
\label{fig6}
\end{figure}
 
\begin{figure}
\caption{
The logarithm of the critical exponents found as a result of a 
simulation for different lattice sizes plotted against $1/n$. 
Estimations of the three exponents in 
the limit of infinitely large lattice size ($1/n\to 0$) are also 
shown. The extrapolated exponents in the limit $n \to \infty$ 
are $\tau_{\infty} = 1.51\pm 0.04$, $\alpha_{\infty} =1.63\pm 0.04$ 
and $\mu_{\infty} = 1.36\pm 0.04$.}
\label{fig7}
\end{figure}
 
\begin{figure}
\caption{
The average avalanche size $<S_T>$ and the average number
of topplings $<m_T>$ as functions of the lifetime $T$,
presented in a double logarithmic plot. The corresponding
slopes are $\beta = 1.13 \pm 0.05$ and $\gamma = 1.75 \pm
0.05$. Data are logarithmically binned at lifetimes $T > 50$.}
\label{fig8}
\end{figure}
 
\begin{figure}
\caption{
Power spectra of the number of particles which drop out of
the system (stars) and of the number of topplings
(diamonds) in a given time step. The data for three system
sizes are depicted, from $n = 5$ to $n = 7$. The results
show $1/f^2$ type of spectra, which indicates an exponential
decay of time correlations. It can be observed that the
frequency at which the $1/f^2$ type of the spectrum crosses
over to the white noise type (i.e. the flat part of the
spectrum), due to the finite size of the system, decreases
with the lattice size. }
\label{fig9}
\end{figure}

\begin{table}
\centering
 \begin{tabular}{||c|c|c|c|c|c||}\hline
$n$ &$p_1$&$p_2$&$p_3$&$p_4$&$\langle z \rangle$\\ \hline
3 & 0.075& .143& 0.306&0.476&3.18\\\hline
4 & 0.070& .136& 0.302&0.492&3.22\\\hline
5 & 0.069& .133& 0.299&0.499&3.23\\\hline
6 & 0.068& .132& 0.298&0.501&3.23\\\hline
7 & 0.068& .132& 0.298&0.502&3.23\\\hline
 \end{tabular}
\caption{The fraction of sites having height equal to $z = 1$,
$2$, $3$ and $4$ and the average height $\langle z \rangle$ as 
found for different system sizes $L = 2^n + 1$
and $n = 3$, $4$, $5$, $6$ and $7$.}
 \end{table}

\end{multicols}
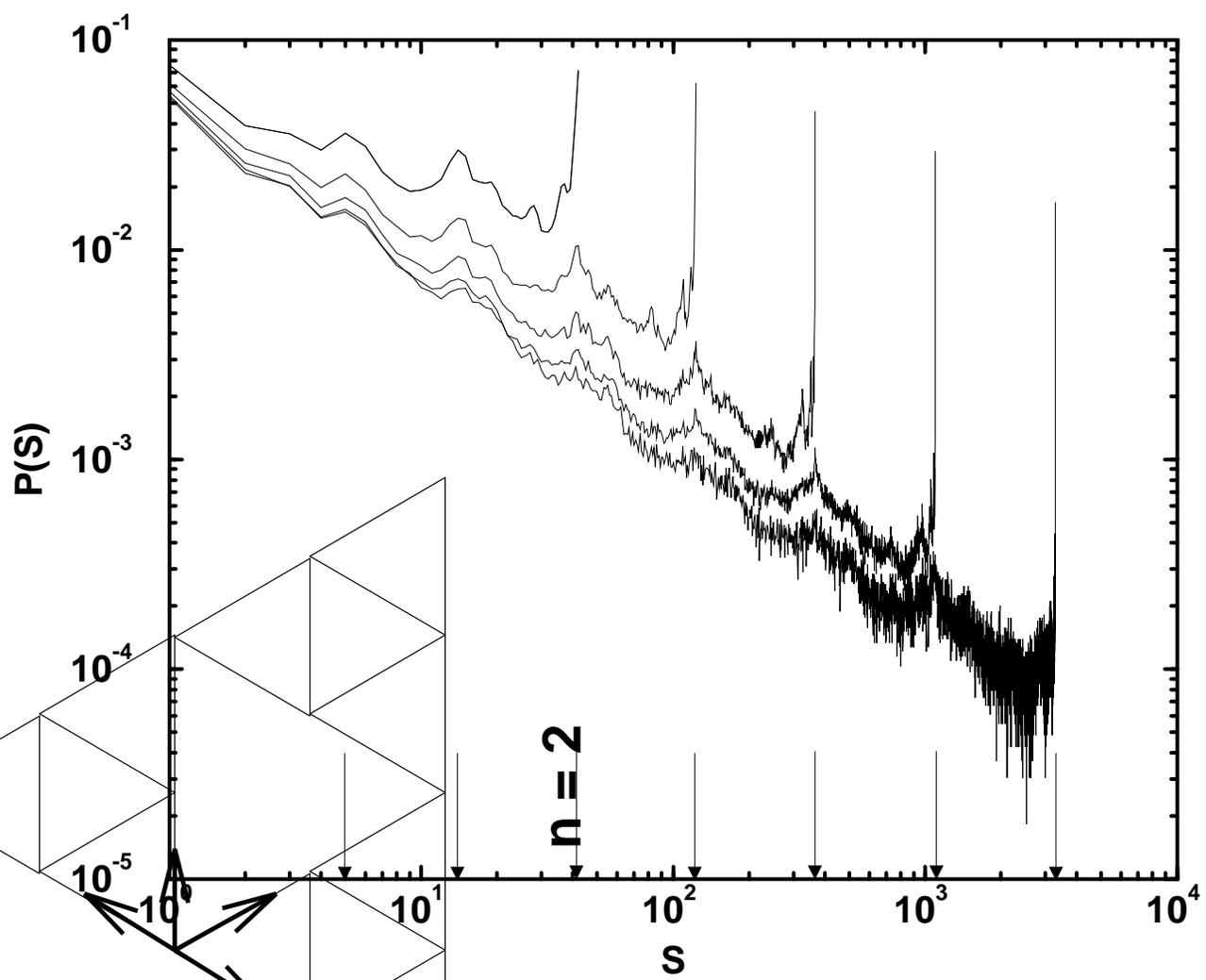
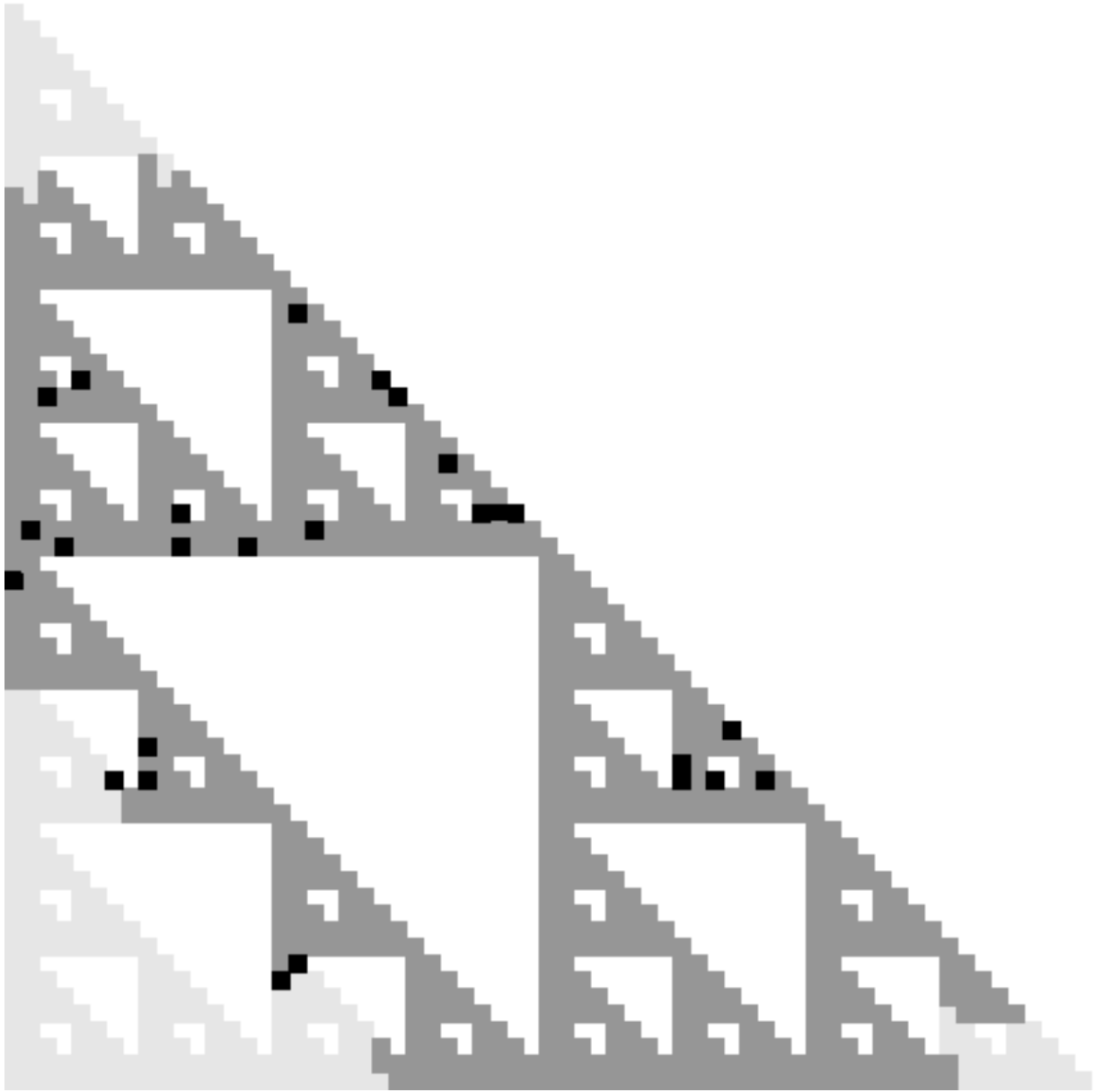
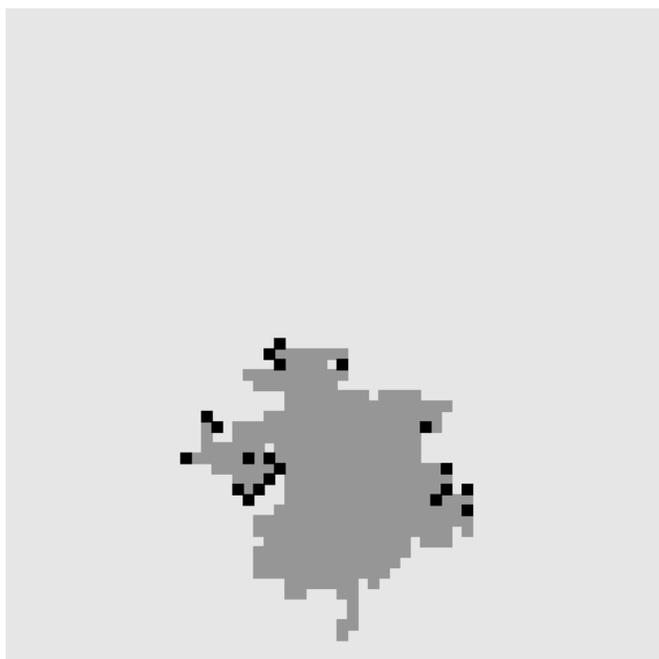
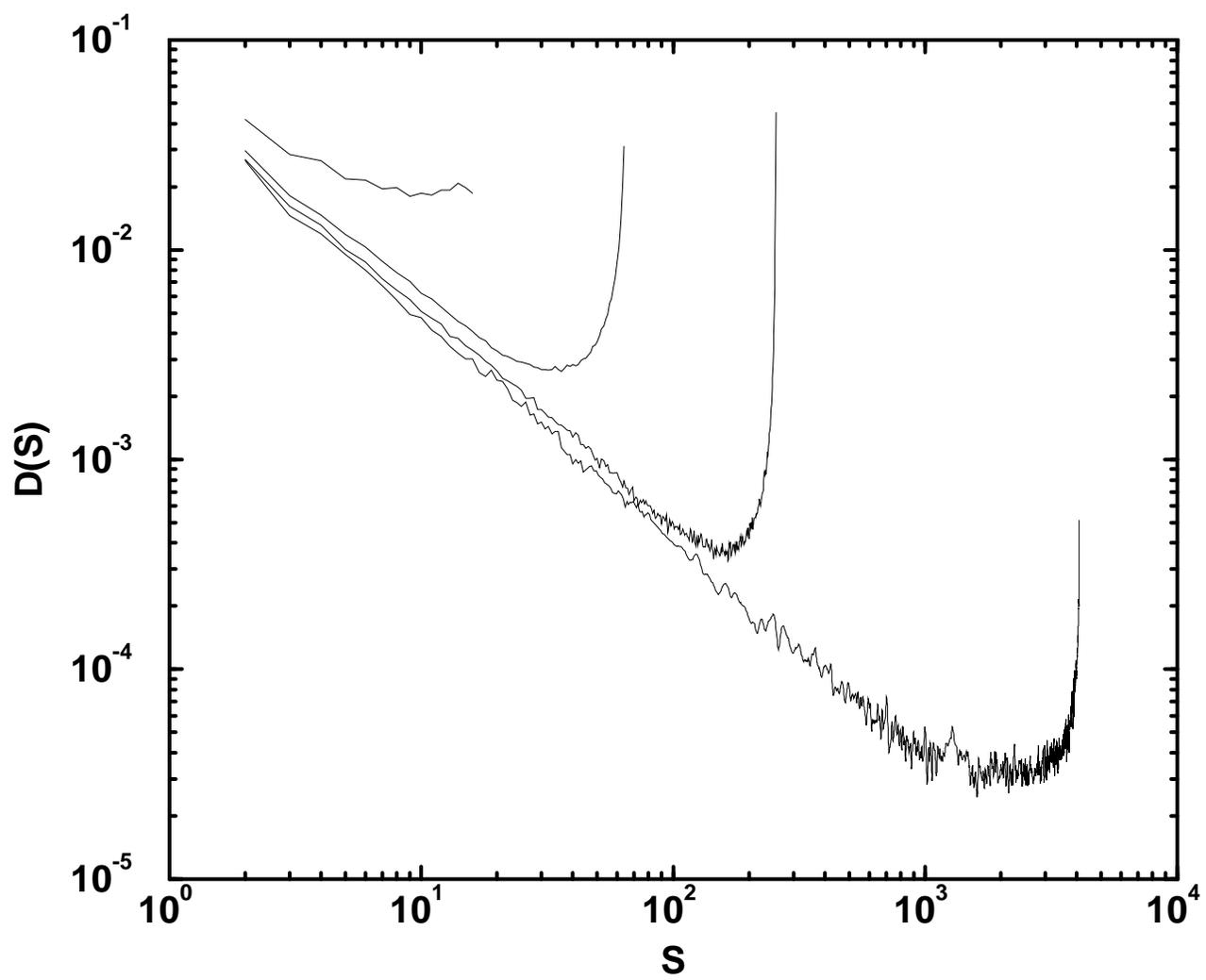
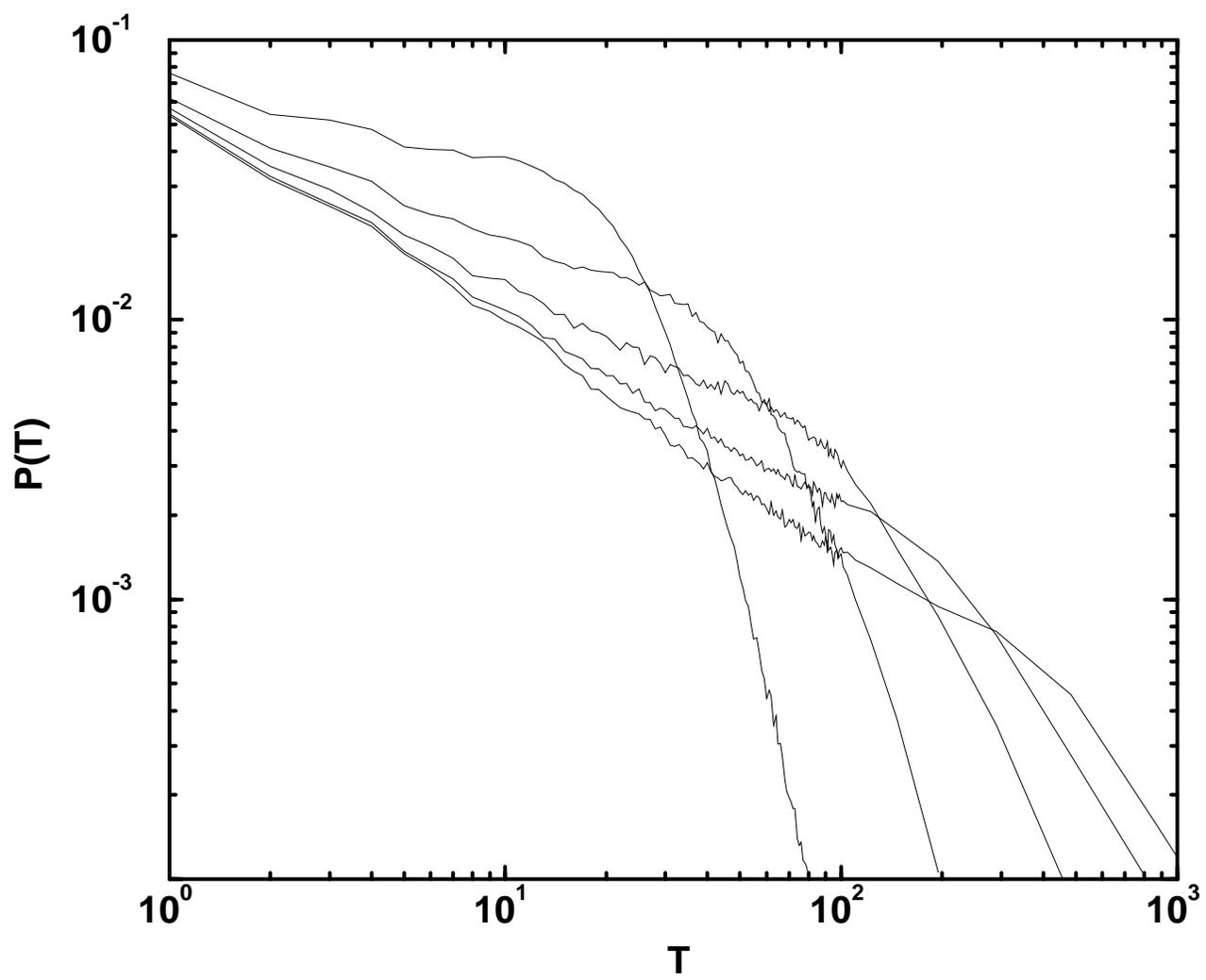
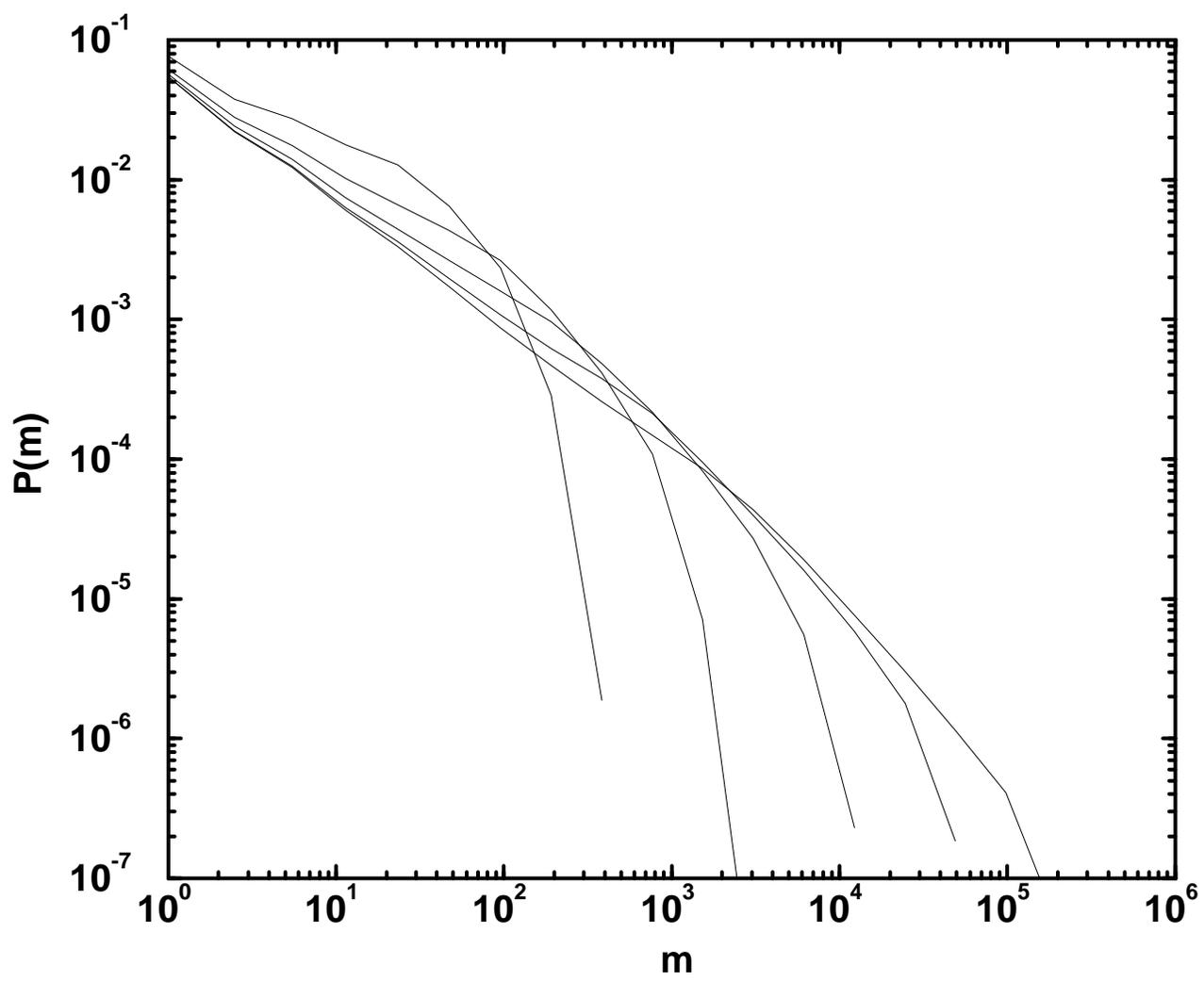
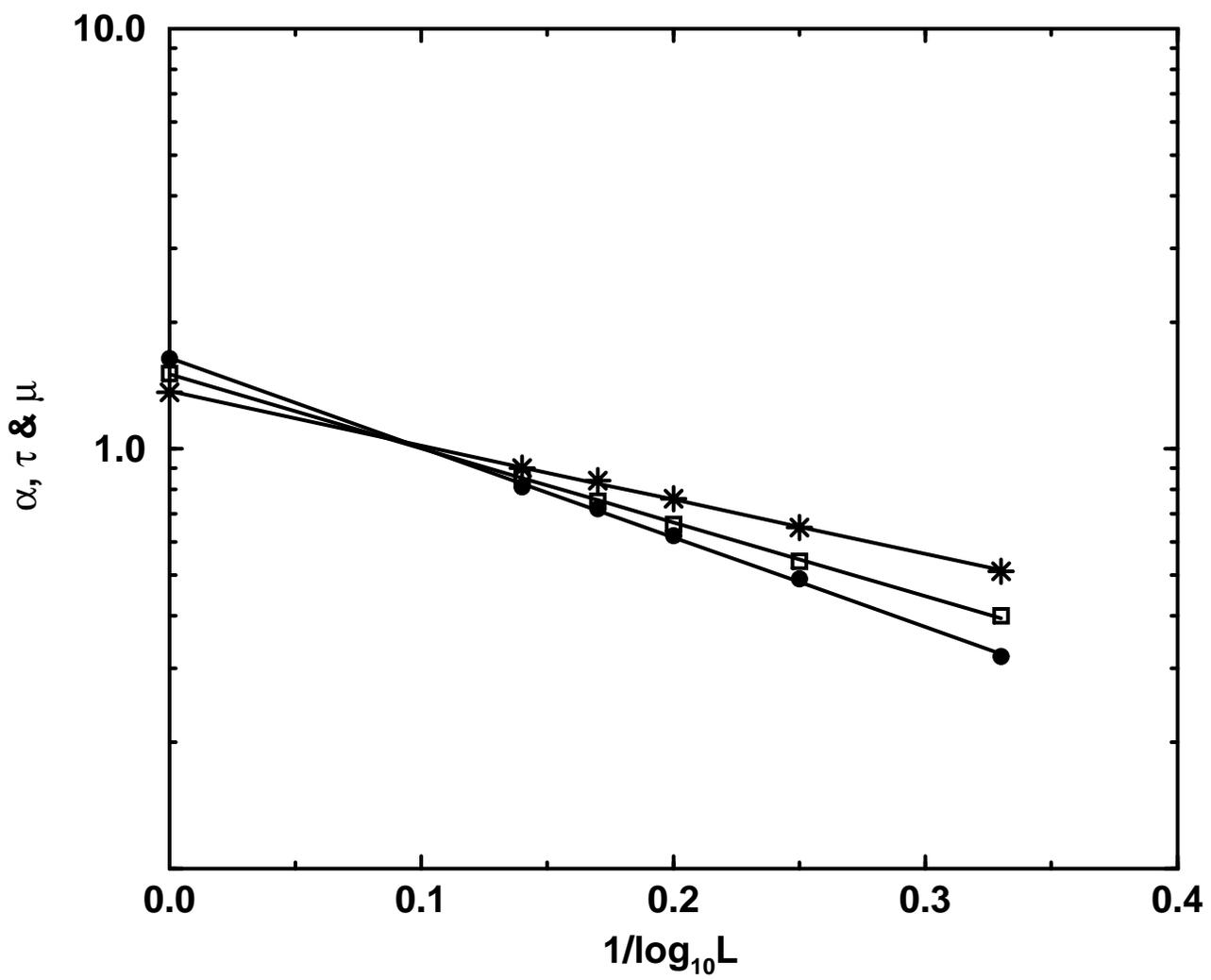
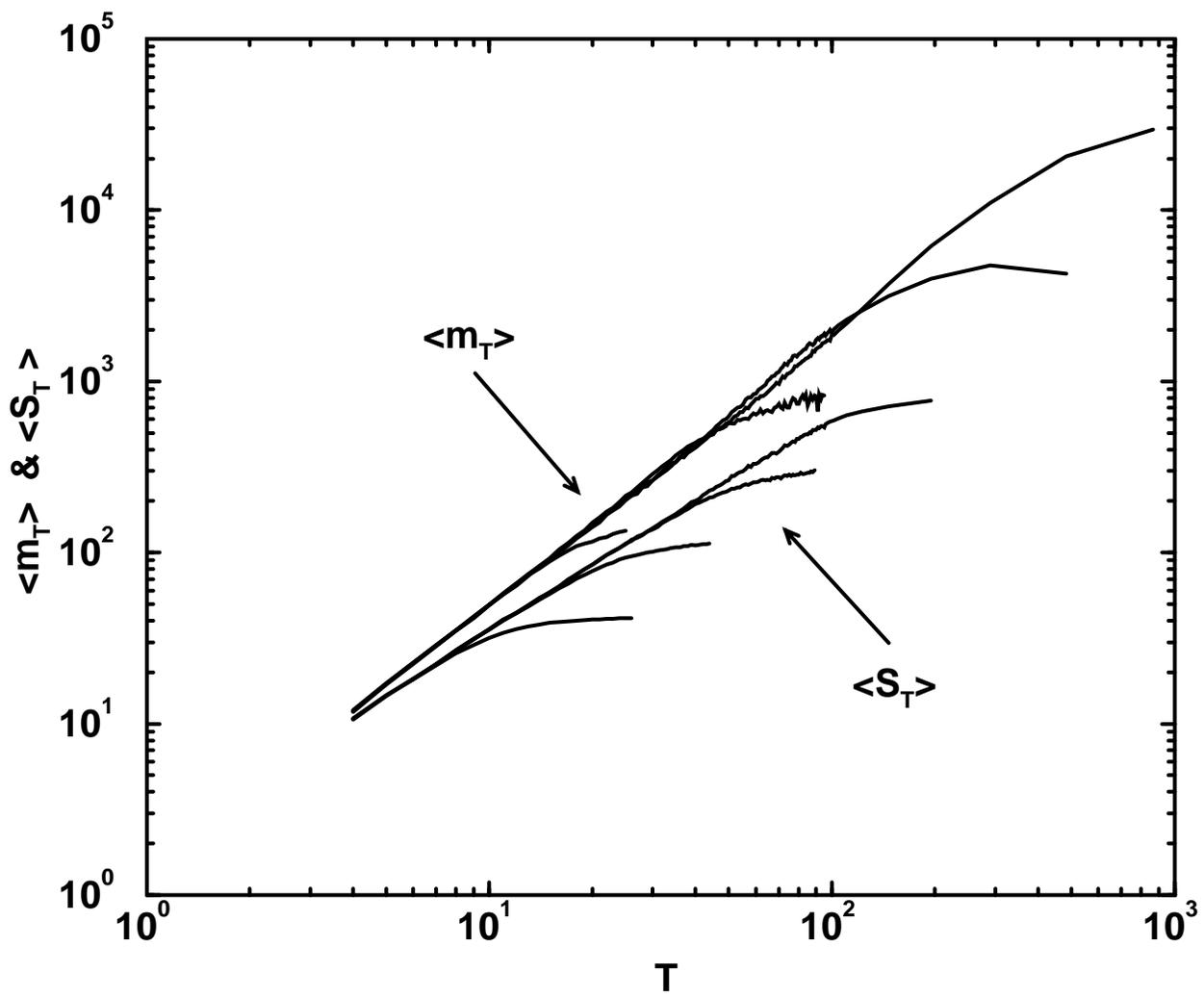
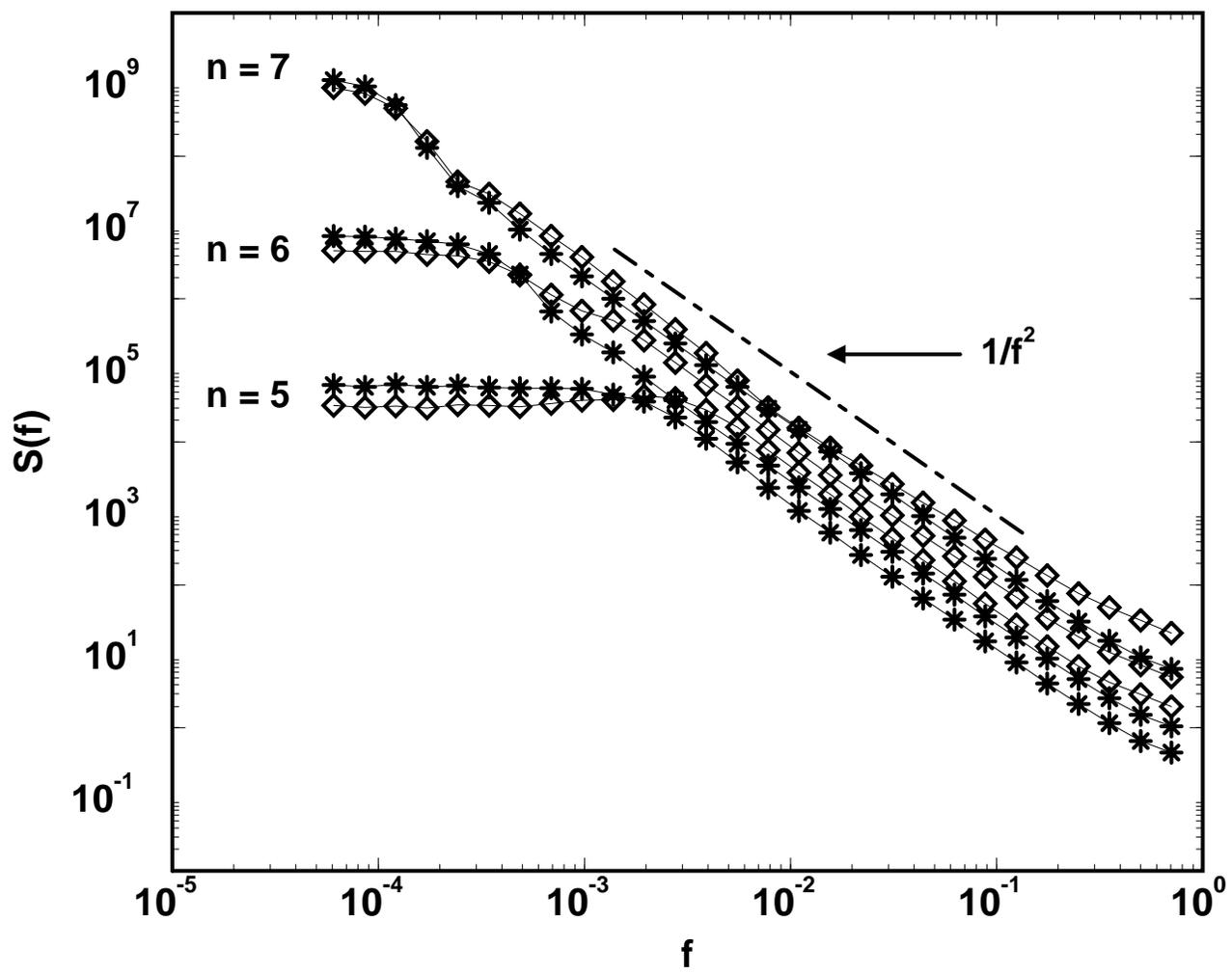
\end{document}